%% file: paper.tex
\documentstyle[epsf,referee]{mn}

\begin{document}
\def\lsun{{\rm L_{\odot}}}
\def\msun{{\rm M_{\odot}}}
\def\rsun{{\rm R_{\odot}}}
\def\be{\begin{equation}}
\def\ee{\end{equation}}


\title{Hydrodynamical winds from two-temperature plasma in X-ray binaries }
\author[David J. Lin, R. Misra and Ronald E. Taam]{ David J. Lin, R. Misra and Ronald E. 
Taam\\ Department of Physics and Astronomy, Northwestern Univ, 2131 
Sheridan Road, Evanston, IL 60208\\ 
d-lin@northwestern.edu, ranjeev@finesse.astro.nwu.edu, taam@apollo.astro.nwu.edu\\}

\newcommand{\lta}{\la}
\newcommand{\gta}{\ga}

\maketitle
\begin{abstract}

Hydrodynamical winds from a spherical two-temperature plasma surrounding a 
compact object are constructed. The mass-loss rate is computed as a function 
of electron temperature, optical depth and luminosity of the sphere, the 
values of which can be constrained by the fitting of the spectral energy 
distributions for known X-ray binary systems. The sensitive dependence of the 
mass loss rate with these parameters leads to the identification of two distinct 
regions in the parameter space separating wind-dominated from non wind 
dominated systems. A critical optical depth ($\tau_c$), as a function
of luminosity and electron temperature, is defined which differentiates these 
two regions. Systems with optical depths significantly smaller than $\tau_c$ 
are wind-dominated. 

The results are applied to black hole candidate X-ray binary systems in the 
hard spectral state (Cyg X-1, GX 339-4 and Nova Muscae), and it is found that 
the inferred optical depth ($\tau$) is similar to $\tau_c$
suggesting that they are wind regulated systems. On the other hand, for X-ray 
binary systems containing a neutron star (e.g., Cyg X-2) $\tau$ is much larger 
than $\tau_c$ indicating the absence of significant hydrodynamical winds.
\end{abstract}

\section{INTRODUCTION}
\label{sect1}

X-ray binary systems containing a black hole candidate are typically found to 
be in two different spectral states. In the hard state, the broad band X-ray 
spectrum can be described as a power-law (photon spectra index $\approx 1.5$) 
with a high energy cutoff around 100 keV. In the soft state, the spectrum 
consists of two components. There is usually an extended power-law (with 
spectral index $\approx 2.5$) and a soft X-ray emission which has a spectral 
shape similar to that of a black body. For a recent review of the observations 
and phenomenological description of these sources see Tanaka \& Shibazaki 
(1996). 

The modeling of the hard state spectra can be described in terms of an 
unsaturated Comptonization process of soft photons in a region with hot 
electrons ($kT \approx 50$ keV) and electron scattering optical depths of 
order unity. In a pioneering study, Shapiro, Lightman \& Eardley (1976) 
identified this hot region with a geometrically thick, optically thin, hot 
accretion disc. In this model, the gravitational energy dissipated in the disc, 
heats the ions which in turn transfer the energy to electrons by Coulomb 
interactions. However, the electron-ion Coulomb interaction rate is inefficient 
in such an environment of low density and high electron temperature, and this 
leads to a large difference in temperature between the electrons and ions, 
with the ion temperature reaching nearly virial values ($10^{11} K$). The 
importance of radial advection of energy in such a disc was noted by Ichimaru 
(1977). Taking advection into account Narayan \& Yi (1994) constructed 
self-similar solutions for the disc equations called Advection Dominated 
Accreting Flows (ADAF) showing not only that the proton temperatures approach
their virial values, but also that the radiative efficiency of accretion can 
be significantly reduced as a result of the advection of energy into the black 
hole. In an alternative description Chakrabarti \& Titarchuk (1995) argued 
that under certain conditions a shock may arise in such accretion discs and 
identified the hot Comptonizing region with the post-shock flow. Despite the 
differences in the geometry, radiative processes and the detailed disc 
structure, both these models have in common the presence of a two-temperature 
plasma. 

Such a plasma is a natural outcome of any accretion disc model which (a) 
identifies the hard state X-ray spectrum as a result of the unsaturated 
Comptonization process of soft photons, (b) assumes that the viscous energy 
dissipated heats the ions and (c) that the only mechanism for energy transfer 
between the ions and electrons is Coulomb interaction. We note here that these 
assumptions may not be valid since the viscous energy dissipated may heat the 
electrons preferentially if a strong equipartition magnetic field is present 
in the disc (Bisnovatyi-Kogan \& Lovelace 2000).  In addition, there could be 
unknown mechanisms which transfer energy between ions and electrons more 
efficiently than the Coulomb interaction. Thus, it will be useful to have an 
independent observational signature which could confirm the existence of 
two-temperature plasmas in black hole candidate systems.

The nearly virial  proton temperature of this plasma suggests the possibility 
of a strong hydrodynamical wind arising from these systems (e.g., Piran 1977;
Takahara, Rosner \& Kusunose 1989; Kusonose 1991). Such an outflow 
could transport away a significant fraction of mass, energy and/or angular 
momentum, thereby affecting the structure and stability of the disc and 
affecting the radiative efficiency of accretion for a given mass transfer rate.
Chakrabarti (1999) and Das (1999) have studied the possibility of outflows in 
the context of the shock/centrifugal barrier models. 
They find that for certain 
values of the disc parameters (e.g., accretion rate and specific entropy) a 
hydrodynamical wind occurs. For the ADAF disc solutions, Blandford \& Begelman 
(1999) argued that only a small fraction ($<$ 1\%) of the gas actually 
falls into the black hole and the rest is driven away as a wind (in an 
advection dominated inflow-outflow solution - ADIOS), thereby effectively 
reducing the radiative efficiency. Beckert (2000) confirmed this result for
different viscosity laws while Quataert \& Narayan (1999) 
showed that the X-ray spectra from such a wind driven accretion process can 
explain the observed spectra of some black hole candidate systems in 
quiescence. These calculations were undertaken for low mass accretion rates, and 
it is not clear how the system will behave in the high accretion rate regime 
inferred for black hole candidate systems in the hard state.

The formation of hydrodynamical winds from accretion discs depends on the 
structure and geometry of the discs. Thus, detailed calculations of the 
outflow are intrinsically model dependent. Further, reliable calculations of 
the structure of such discs are difficult because uncertainties exist 
in the vertical distribution of energy dissipation in the disc associated with 
our lack of detailed understanding of the viscosity.  We note that internal 
magnetic fields in the disc could also facilitate (or inhibit) the formation 
of winds. Here, electromagnetic forces may accelerate and collimate the wind 
to form high velocity jets as observed in microquasar systems (Mirabel \& 
Rodriguez 1999).  Considering these uncertainties a prudent approach would be 
to estimate the wind characteristics using only those parameters which can 
be directly constrained by the fitting of spectral energy distributions.
Such an analysis would allow a rough estimation of the magnitude of the mass and energy 
lost in the form of a wind  for a system in question.  
With such an objective in 
mind, we report in this paper on calculations of the structure of an 
hydrodynamical wind 
similar to those found by Takahara et al (1989), but in 
the context of a uniform spherical two-temperature plasma 
around an accreting black hole. Rather than treating the entire disc/wind 
configuration with detailed heating and cooling processes as in Kusunose (1991) 
the calculations are parameterized in terms of 
the electron temperature of the cloud, $T_e$, the optical depth, $\tau$, and 
the total luminosity of the source.  We adopt this approach as these parameters 
are constrained by spectral fitting analyses. 
In the next section, the formulation of the problem and the numerical results 
are presented.  The application of these results to observed systems is given 
in \S 3 and discussed in the last section.  

\section{Hydrodynamical winds}
\label{sect2}

The mass loss rate due to a hydrodynamical wind depends 
sensitively on the ion temperature. This allows a system to be defined as 
wind-dominated if the ion temperature is greater than some critical value 
($T_c$), while for temperature less than this value the mass and energy loss 
rate is small enough not to affect the dynamics of the system. This critical 
temperature will scale as the virial one i.e. $ kT_c \propto GM/R$ where $R$ 
is the size of the region and $M$ is the mass of the black hole. For a 
two-temperature plasma the power output is mediated by 
electron-proton Coulomb interactions and the luminosity, L, is given by 

\begin{equation}
L/V  =  {3\over 2} n (kT_i - kT_e) \nu_{ep}
\end{equation}
where $ V = 4/3 \pi R^3$ is the volume of the sphere, $n$ is the average number 
density, $T_i$ and $T_e$ are the ion and electron temperatures, and $\nu_{ep}$ 
is the frequency of the electron-ion Coulomb interaction.  A critical optical 
depth $\tau_c (=  n \sigma_T R)$ may be defined for the system, wherein for 
systems with $\tau < \tau_c$, the ion temperature is larger than $T_c$ 
(see below).  Taking 
the above relationships into account with the Coulomb exchange rate given as 
(Spitzer 1962)
\begin{equation}
\nu_{ep} = 2.4 \times 10^{21} ln \Lambda \rho T_e^{-3/2}
\end{equation}
where $ln \Lambda (\approx 15)$ is the Coulomb logarithm and $\rho$ is the 
mass density, it follows that 
\begin{equation}
\label{tauest}
\tau_c \propto L^{1/2} T_e^{3/4} M^{-1/2} 
\end{equation} 
Since $L$ and $T_e$ are parameters which can be constrained by spectral 
fitting, $\tau_c$ can be directly estimated for a black hole of a given mass. 
Spectral fitting can also give information about the optical depth of the 
system, which can then be compared with $\tau_c$, to determine whether the 
system could be wind-dominated or not. The above analysis shows that $\tau_c$ 
does not depend on the size of the system $R$. This is fortunate since $R$ is 
not well constrained by observations.

To quantify the above analysis we solve for the hydrodynamical wind structure 
from a spherical two-temperature plasma with an optical depth ($\tau$), 
electron temperature $T_e$ and luminosity $L$. The central mass is taken to 
$M = 10 M_\odot$ and the radius of the sphere is fixed at $R = 20 GM/c^2$. We 
use the basic equations of the hydrodynamical theory of stellar winds (Parker 
1958) which are the conservation of radial momentum,
\begin{equation}
\rho v {d v \over dr} = -{dP \over dr} - \rho {G M \over r^2}
\end{equation}
and the conservation of mass, i.e. the mass outflow 
\begin{equation}
\dot M_o = \rho v 4 \pi r^2 = constant.
\end{equation}
Here $v$ is the radial velocity, $P = K \rho^\Gamma$ is the pressure and 
$\Gamma$ is the adiabatic index. These equations can be combined to give,
\begin{equation}
{d v \over d r} ={ [ {2 c_s^2 \over r} - {G M \over r^2}] \over [ v - 
{ c_s^2 \over v}]}
\end{equation}
where $c_s = (dP/d\rho)^{1/2}$ is the sound speed. At the sonic point, both 
the numerator and denominator vanish. Wind solutions were constructed by 
integrating from the sonic point. The inner boundary is taken at $R$ where
the number density $n = \tau/(R \sigma_T)$ and ion temperature is calculated
using equation (1). The solution is constructed such that at infinity both
the density and pressure tend to zero. A wind solution does not exist if
the ion temperature at the surface 
$kT_i < kT_{min} \approx  ({\Gamma-1 \over \Gamma}) G M m_p /R$. For
high values of $kT_i > kT_{max} \approx ({\Gamma+1\over 4\Gamma })G M m_p/R$, the flow is supersonic
at the surface i.e. the sonic point is actually located at a radius less
than $R$. Thus we restrict this analysis only to those values of $\Gamma$
such that $kT_{min} < kT_i < kT_{max}$. Note that for $\Gamma =  5/3$, 
$kT_{min} \approx kT_{max}$ and there does not exist a hydrodynamical wind 
solution with a sonic radius greater than $R$ for any $kT_i$. Thus this 
analysis is restricted to values of $\Gamma < 5/3$. 
Such regimes
have also been studied by Chakrabarti (1999) and for example, in the limit of 
efficient conduction of heat an isothermal wind flow results 
corresponding to $\Gamma \approx 1$. Radiative processes can also
lead to redistrubution of energy in the wind leading to $\Gamma < 5/3$
(Chakrabarti 1999).

In Figure 1, the calculated mass loss rate as a function of ion temperature 
is illustrated for a fixed optical depth ($\tau = 1$). The virial temperature
$T_v \equiv G M m_p/ k R$ for the sphere is $ \approx 5 \times 10^{11}$ K.  
As expected, $\dot M_o$ 
is sensitive to the ion temperature: a factor of two increase in $T_i$ causes 
an increase of $\dot M_o$ by two orders of magnitude. 
This justifies defining a 
critical temperature beyond which the system is wind-dominated and below
which it is not. 
For comparison, we note that the ion temperatures in ADAF type 
solutions correspond to $T_i \sim  T_v$, 
which in the presence of winds are lowered (Misra \& Taam 2000). 
The mass loss  
rate can be expressed in terms of luminosity, optical depth, and electron temperature.
For a typical luminosity ($L = 10^{38}$ ergs/s) and electron temperature 
($T_e = 50$ keV), the variation of mass loss rate with optical depth is shown 
in Figure 2. Here, the sensitivity of $\dot M_o$ with $\tau$ is apparent: a 
factor of two increase in $\tau$ decreases $\dot M_o$ by at least 
three orders of 
magnitude. We show the contour plots of $\dot M_o$ for the $\tau$, $T_e$  
and $\tau$, $L$ planes in Figures 3 and 4 respectively.  These plots 
highlight the steep variation of $\dot M_o$ with all three parameters. 

The $\tau$, $T_e$ and $L$ parameter space can then be divided into two regions 
corresponding to high and low values of $\dot M_o$. In particular we define 
the wind dominated region as corresponding to $\dot M_o > \dot M_{crit} $ and 
the complementary region as the non wind dominated one. The choice of $\dot 
M_{crit}$ is ad-hoc and is chosen to be $10^{18}$ g/s. Since the dependence of 
$\dot M_o$ on the parameters is steep, choosing $\dot M_{crit}$ to be $10^{17}$ 
g/s still leads to a division of the parameters space to nearly similar 
regions.  A critical optical depth $\tau_c$ can be defined as a function of $L$ 
and $T_e$ for which $\dot M_o = 10^{18}$ g/s.  The topology of Figures 3 and 
4 and the qualitative analysis described in the beginning of the section 
(eqn 3) suggests that $\tau_c$ can be represented as:
\begin{equation}
\label{tauc}
\tau_c = A ( {k T_e \over 50 \;\hbox {keV}})^\alpha ( {L \over 10^{38} \; \hbox {ergs/s}})^\beta
\end{equation}
We constrain $A$, $\alpha$ and $\beta$ for different values of $\dot M_{crit}$, 
$\Gamma$, $M$, and $R$ and present the result in Table 1.


\begin{table*}
\begin{minipage}{\hsize}
\caption{Values of $A$, $\beta$ and $\alpha$  in equation (\ref{tauc}). Here $\dot M_o$ is in
units of g/s and $R$ is in units of $r_s = 2 GM/c^2$.}

\begin{tabular}{ccccccc}\hline
log $\dot M_{crit}$ &  $\Gamma $ & $M (M_\odot)$ &$ R (r_s)$& A & $\alpha $ & $\beta$ \\ \hline
 & & & & & & \\
 18 & 1.05 & 10 & 10 & 1.9 & 0.90 & 0.60 \\
 17 & 1.05 & 10 & 10 & 2.5 & 0.84 & 0.56 \\
 18 & 1.2  & 10 & 10 & 1.7 & 0.85 & 0.56 \\
 18 & 1.05 & 30 & 10 & 1.3 & 0.86 & 0.57 \\
 18 & 1.05 & 10 & 100& 3.7 & 0.90 & 0.58 \\
\hline

\end{tabular}
\end{minipage}
\end{table*}

From these results it can be seen that $A$, $\alpha$ and $\beta$ do not vary 
greatly with changes in $\dot M_{crit}$, $\Gamma$, $M$ or $R$. Thus to within 
a factor of two, the critical optical depth can be represented as
\begin{equation}
\label{taucn}
\tau_c = 2 ( {k T_e \over 50 \;\hbox {keV}})^{0.85} ( {L \over 10^{38} \; \hbox {ergs/s}})^{0.55}
\end{equation}
We can now use equation (\ref{taucn}) to determine from observations if a 
particular system is wind dominated or not.

\section{Application to X-ray binaries}
\label{sect3}

Assuming that a two-temperature plasma configuration is established subject 
to the conditions described in \S 2, we discuss the possible applications of 
hydrodynamical winds to binary systems containing compact objects.  In the 
following we first discuss the results for the black hole candidate systems 
Cyg X-1, GX339-4, and Nova Muscae.  For convenience, the results are summarized 
in Table 2. 


\begin{table*}
\begin{minipage}{\hsize}
\caption{Comparison of inferred and critical optical depth for various
X-ray binaries. The luminosity, electron temperature ($T_e$) and
optical depth ($\tau$) have been quoted from the references. The symbols
WD,WR and NW stand for wind dominated, wind regulated and no wind respectively. 
For GS 1124-68 (Nova Muscae) and Cygnus X-1 (soft state) the
electron temperature is assumed and the optical depth has been estimated using
equation (\ref{alphaS}). Note that for the soft state, the assumptions used
in this analysis may not be valid (see text). }

\begin{tabular}{ccccccc}\hline
Source Name & Luminosity (ergs/s)  & $k T_e$ (keV) &$ \tau $& $\tau_c$ & Comment & ref. \\ \hline
Black Holes & & & & & & \\ \hline
& Hard State & & & & &\\ \hline
 & & & & & & \\
 Cygnus X-1  & $3.5 \times 10^{37}$ & 140  & 1 & 2.6 & WR & 1 \\
 GX 339-4 & $3 \times 10^{37}$  & 48 & 1.9  & 3.6 & WR & 2 \\
GS 1124-68 & $4.3 \times 10^{36}$ & 100 & 1.7 & 0.6 & NW & 3 \\
GS 1124-68 & $4.3 \times 10^{36}$ & 200 & 1.2 & 1.1 & WR & 3 \\ \hline
& Soft State  & & & & &\\ \hline
Cygnus X-1  & $2.9 \times 10^{37}$ & 200  & 0.40  & 3.3 & WD & 4 \\
GS 1124-68  & $2.5 \times 10^{37}$ & 200  & 0.37  & 3.0 & WD & 3 \\ \hline
Neutron Stars & & & & & & \\ \hline
& Low luminosity & & & & &\\ \hline
XB 1608-52  & $4.1 \times 10^{36}$  & 7 & 7.6  & 0.13 & NW & 5 \\
XB 1636-536  & $2.7 \times 10^{37}$  & 1.8 & 16.7  & 0.12 & NW & 5 \\
XB 0748-676  & $3.7 \times 10^{36}$  & 2.4 & 24.1  & 0.04 & NW & 5 \\
XB 1254-69  & $7.8 \times 10^{36}$  & 1.9 & 15.6 & 0.06 & NW & 5 \\
XB 1820-30  & $1.4 \times 10^{37}$  & 3.5 & 13.2& 0.15 & NW & 5 \\ \hline
 & High luminosity & & & & &\\ \hline
XB 1820-30  & $4.7 \times 10^{38}$  & 3.3 & 11.9& 1.2 & NW & 5 \\ 
Cyg X-2  & $1 \times 10^{38}$  & 3.7 & 9.4& 0.5 & NW & 5 \\ 
GX 17+2  & $2.5 \times 10^{38}$  & 3 & 13 & 0.8 & NW & 5 \\ 
GX 9+1  & $1.7 \times 10^{38}$  & 3 & 11 & 0.6 & NW & 5 \\ 
GX 349+2  & $2.2 \times 10^{38}$  & 3.7 & 10 & 0.9 & NW & 5 \\ 
\hline
 
\end{tabular}

References: 1: Gierlinski et al. (1997), 2: Zdziarski et al. (1998), 3: Ebisawa et al. (1994), 4: Gierlinski et al. (1999), 5: White,Stella \& Parmar (1988).
\end{minipage}
\end{table*}

The hard state spectrum of Cyg X-1 has been fitted by Gierlinski et al. (1997) 
with a Comptonization model, and they obtained the following parameters: $\tau 
\approx 1, kT_e \approx 140$ keV and $L = 3.5 \times 10^{37}$ ergs/s (for a 
distance of 2.5 kpc). Using these values we find from equation (\ref{taucn}) 
for a black hole of $10 M_{\odot}$ that $\tau_c = 2.6$ which is on the order of
the observed $\tau \approx 1$. Similarly for another black hole system, GX 
339-4 in the low state, Zdziarski 
(1998) obtained $\tau = 1.93$, $kT_e = 48$ keV and $L = 3 \times 10^{37}$ 
ergs/s. For these values $\tau_c = 3.6$ which is again of the order of the 
observed $\tau$. Thus, our analysis of both Cygnus X-1 and GX 339-4 in their 
hard states indicates that these systems are on the borderline of having a 
strong or weak wind. This may point to the existence of a feedback mechanism 
wherein the energy and mass loss from the wind regulate the hot disc 
structure. 

Unlike the persistent systems discussed above, black hole X-ray novae are
transients. In these sources, the luminosity increases by several orders of 
magnitude in $\approx 10$ days and then decays exponentially in a time-scale
of $\approx 1$ month. During the peak of the outburst the system is usually 
in the soft spectral state and makes a transition to the hard state in $\approx 
2-3$ months (see Tanaka \& Shibazaki 1996). Observations of the X-ray transient 
source Nova Muscae (GS 1124-68) by the Ginga satellite through most of its
evolution were reported in Ebisawa et al. (1994). Since the Ginga energy band 
is restricted to $< 20$ keV, the roll-over of the spectrum at high energies 
was not observed. Thus, a Comptonization fit does not constrain both $T_e$ and 
$\tau$. In particular only a low energy spectral index ($\alpha_s$) (during
different times of the evolution) was measured. For a Comptonized spectrum, 
$\alpha_s$ is approximately related to $T_e$ and $\tau$ by,
\begin{equation}
\label{alphaS}
-(3+\alpha_s) = -{3\over2} \pm \sqrt{{9\over 4}+{4 \over y}}
\end{equation}
where, $y = (4 k T_e/m_ec^2)$ max$(\tau,\tau^2)$ is the Compton y-parameter. 
For Nova Muscae in the hard state, $\alpha_s$ was observed to $\approx 0.5$ and 
at a luminosity, $L = 4.3 \times 10^{36}$ ergs/s (Ebisawa et al. 1994). If we 
now assume that $T_e \approx 100$ keV one obtains from equation (\ref{alphaS}), 
$\tau \approx 1.7$. The critical optical depth for such a system is then 
$\tau_c = 0.6$ which is less than the inferred $\tau$, indicating the absence 
of a strong wind. However, a strong wind is indicated if one assumes $k T_e 
\approx 200$ keV instead, since the inferred $\tau \approx 1.2$ which is close 
to the critical value of $\tau_c = 1.1$.  Thus, hydrodynamical winds can be 
important in the hard state of Nova Muscae provided that the electron 
temperatures are $\ga 150$ keV.  

In the soft state of black hole candidate 
systems the power-law spectrum is steeper than the 
hard state. Gierlinski et al. (1997) analyzed the soft state Cygnus X-1 data and found that
the energy spectral index $\alpha_s \approx 1.5$ and the power-law extends up to 
$\approx 200$ keV with no apparent cutoff. They argued 
that thermal Comptonization does not describe the spectra well. Instead, the 
power-law is probably due to non-thermal Comptonization. In this case, the 
presence of 
non-thermal electrons indicates that the electrons are probably heated directly 
instead of mediated by processes involving protons. Thus, the basic assumption 
underlying this study is probably not valid for the soft spectral state of 
Cygnus X-1. Nevertheless, 
from the observed $\alpha_s = 1.4$, $L = 3 \times 10^{37}$ 
ergs/s and assuming that $kT_e \approx 200$ keV, one would infer using equation 
({\ref{alphaS}}) that $\tau \approx 0.4$. This is much smaller than the critical 
value of $\tau_c = 3.3$, perhaps indicating the presence of a strong wind.
A similar result is obtained from the soft state of Nova Muscae (Table 2). 

In addition to black hole candidates, the spectra from X-ray binary systems 
containing a neutron star can also be described as being due to Comptonization.
Hot two temperature accretion flows around neutron stars have been recently
constructed by Medvedev \& Narayan (2000).
On the other hand, the hot region may not be the accretion disc itself 
but could be instead an extended
corona surrounding the boundary layer between the disc and the surface of the 
star. This analysis can be still applied to these systems as long as the 
assumption that the total luminosity of the source is channeled by the protons 
to the electrons remains valid.  The critical optical depth when $M = 1.4 
M_\odot$ and $R = 10$ kms is found to be
\begin{equation}
\label{taucn1}
\tau_c = 6 ( {k T_e \over 50 \;\hbox {keV}})^{0.95} ( {L \over 10^{38} \; 
\hbox {ergs/s}})^{0.62}
\end{equation}
Spectral fitting of EXOSAT data from several neutron star binaries has been 
undertaken by White, Stella \& Parmar (1988) which are summarized in Table 
2. For Cygnus X-2, they constrain $kT_e \approx 3.7$ keV, $\tau = 9.4$ and 
$L = 10^{38}$ ergs/s (for a distance of 8 kpc). Using equation (\ref{taucn1}) 
we find that $\tau_c = 0.5$ which is significantly less than the observed 
value. Similar results were obtained for other neutron star systems in both
in the high and low luminosity levels (Table 2). 
This implies that unlike black hole
candidate X-ray binary systems, their neutron star counterparts are not  
dominated by a hydrodynamic wind. This result is further supported by the
analysis undertaken by  Medvedev \& Narayan (2000) who constructed 
self-similar accretion flows onto neutron stars and found that 
hydrodynamic winds are not important in these systems.

\section{Summary and Discussion}
\label{sect4}

The possible occurrence of hydrodynamical winds in X-ray binary systems has  
been investigated.  From simple considerations, the conditions under which 
these winds can be important have been identified.  It is found that  
the mass loss rate in such winds depends sensitively on the luminosity of the
source as well as the electron temperature and optical depth of the coronal 
region.  The steep dependence of the mass loss rate on these parameters
facilitates the use of a critical optical depth to indicate whether a given 
system can support such a wind. Application of the theory indicates that 
winds can exist in the hard state of the black hole candidates Cyg X-1, GX 339-4 
and GS 1124-68.  

Strong winds in these systems may decrease the radiative efficiency
($\eta = L/\dot M_i c^2$, where $\dot M_i$ is the mass inflow rate) by carrying
away a substantial amount of matter and energy. For Nova Muscae, the
radiative efficiency has been estimated to be $\eta = 0.01$ for the soft state
and $\eta = 0.05$ for the hard state (Misra 1999). 
These values are lower than that expected from an Keplerian disc ( $\eta \approx 0.1$)
which could be due to the presence of a strong wind rather
than energy advection into the black hole. Winds may also effect the thermal
stability of accretion discs by introducing an additional channel for energy
loss (e.g., Piran 1977).

We reiterate that the analysis undertaken in this paper 
is based on the assumption that the 
gravitational energy dissipated in the system is transferred to the electrons 
by the ions via Coulomb interactions. This naturally restricts the analysis to 
only certain systems where this is valid. As mentioned earlier, the soft 
state spectrum of black hole candidate systems is probably of a non-thermal 
origin and, hence, the analysis performed in this study cannot be applied. 
Even for the hard spectral states of compact X-ray binaries considered here, 
alternate models to a two -temperature plasma description have been proposed. 
For example, the X-ray spectra could be due to magnetic flare activity above 
a cold disc (Poutanen \& Fabian 1999) or produced by a disc with a rapidly 
varying radial temperature profile (Misra, Chitnis  \& Melia 1998). Thus, 
it should be emphasized that the results presented here are specifically 
discussed within the framework of a two-temperature plasma model. 

The spherical geometry assumed here is simplistic even though two-temperature
discs are geometrically thick. 
Further the effect of angular momentum or of
convection on the dynamics of accretion and the wind outflow have not been 
taken into account. The former effect can  
increase the mass outflow rate as a result of centrifugal support, whereas 
the latter effect can affect the thermal structure of the underlying disk and, hence, 
the existence of a wind.  In this context, the thermal structure is dependent 
on the direction in which angular momentum is radially transported
by convection and the magnitude of the viscosity parameter (see Narayan, 
Igumenshchev, \& Abramowicz 2000).  
A more detailed analysis should take these 
effects and radiative heating/cooling of the wind into account
in at least two spatial 
dimensions (e.g., 
Chakrabarti \& Molteni 1993). 
However, such analyses will, by necessity, be 
model dependent and limited by the uncertainties in the vertical structure and 
geometry of the hot disc. 

The hydrodynamical winds described in this paper may be confined to form a 
jet-like structure by the geometry of the disc and magnetic fields. Further if 
they are accelerated to relativistic speeds by electromagnetic forces, they may 
provide the origin of the radio jets observed in the black hole candidate 
systems known as microquasars (Mirabel \& Rodriguez 1998).  On the other hand, 
the radio jets may be a different phenomenon unrelated to hydrodynamical winds.
In that case it is desirable to have direct observational signatures of these 
outflows. Since the Thomson optical depth of the winds calculated here 
(i.e. the optical depth from the surface of the sphere $R$ to infinity) is 
typically $< 0.1$, column densities of the order of $10^{23}$ cm$^{-2}$ are 
indicated. Although the X-ray continuum spectra are not expected to be altered 
by the outflow, a significant fraction of the column density may not be highly 
ionised giving rise to observable absorption and/or emission lines. 
The detection of P-Cygni type profiles by high resolution X-ray satellites  
Chandra and Newton-XMM may provide evidence for the existence of such winds 
and serve as a useful 
diagnostic not only of the wind structure, but also the 
physics underlying the disc as well. 

\bigskip
RM acknowledges support from the Lindheimer Fellowship at 
Northwestern University.

\newpage

\input psbox.tex
\newpage
\begin{figure*}
\hspace{-1.5cm} 
{\mbox{\psboxto(17cm;20cm){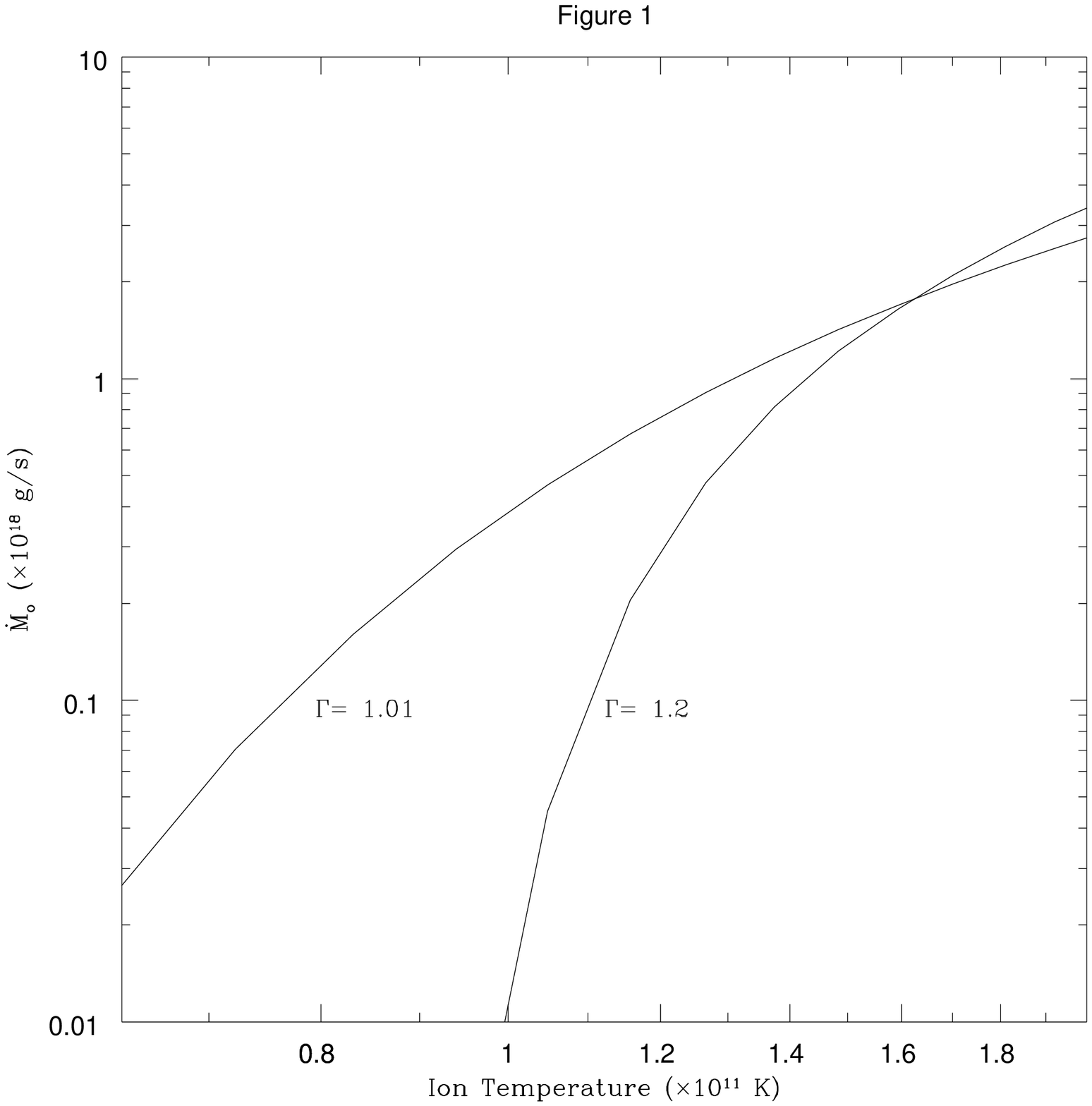}}}
\caption{\label{fig:1}The variation of the mass outflow rate with ion temperature for adiabatic
indices $\Gamma = 1.01$ (nearly isothermal) and $\Gamma = 1.2$. The other parameters
are: $M = 10 M_\odot$, $R = 20 GM/c^2$ and $\tau = n\sigma_T R = 1$.}
\end{figure*}

\begin{figure*}
\hspace{-1.5cm} 
{\mbox{\psboxto(17cm;20cm){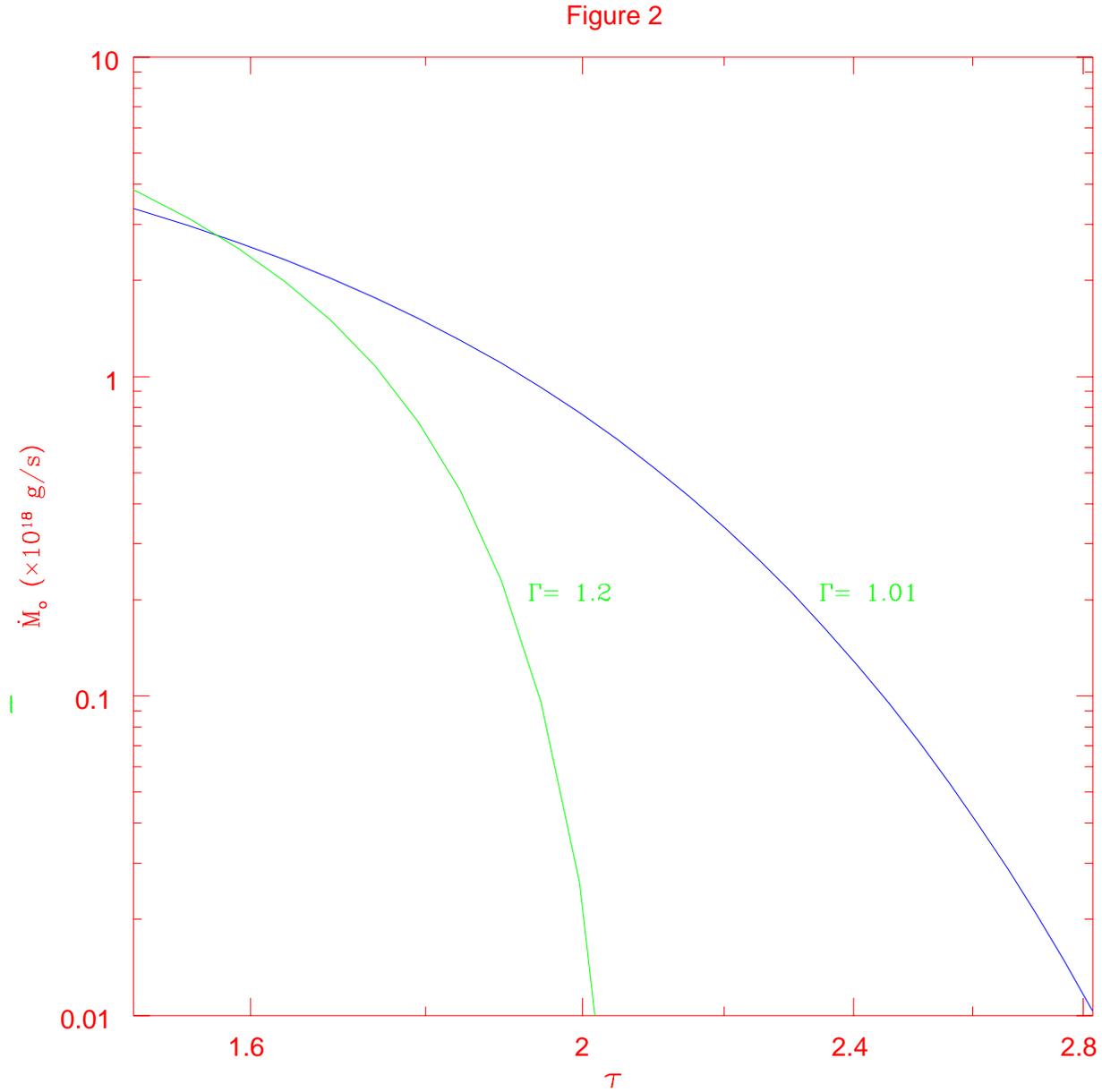}}}
\caption{\label{fig:2}The variation of the mass outflow rate with optical depth for adiabatic
indices $\Gamma = 1.01$ (nearly isothermal) and $\Gamma = 1.2$. The other parameters
are: $M = 10 M_\odot$, $R = 20 GM/c^2$, $kT_e = 50$ keV and $ L = 10^{38}$ ergs/s}
\end{figure*}

\begin{figure*}
\hspace{-1.5cm} 
{\mbox{\psboxto(17cm;20cm){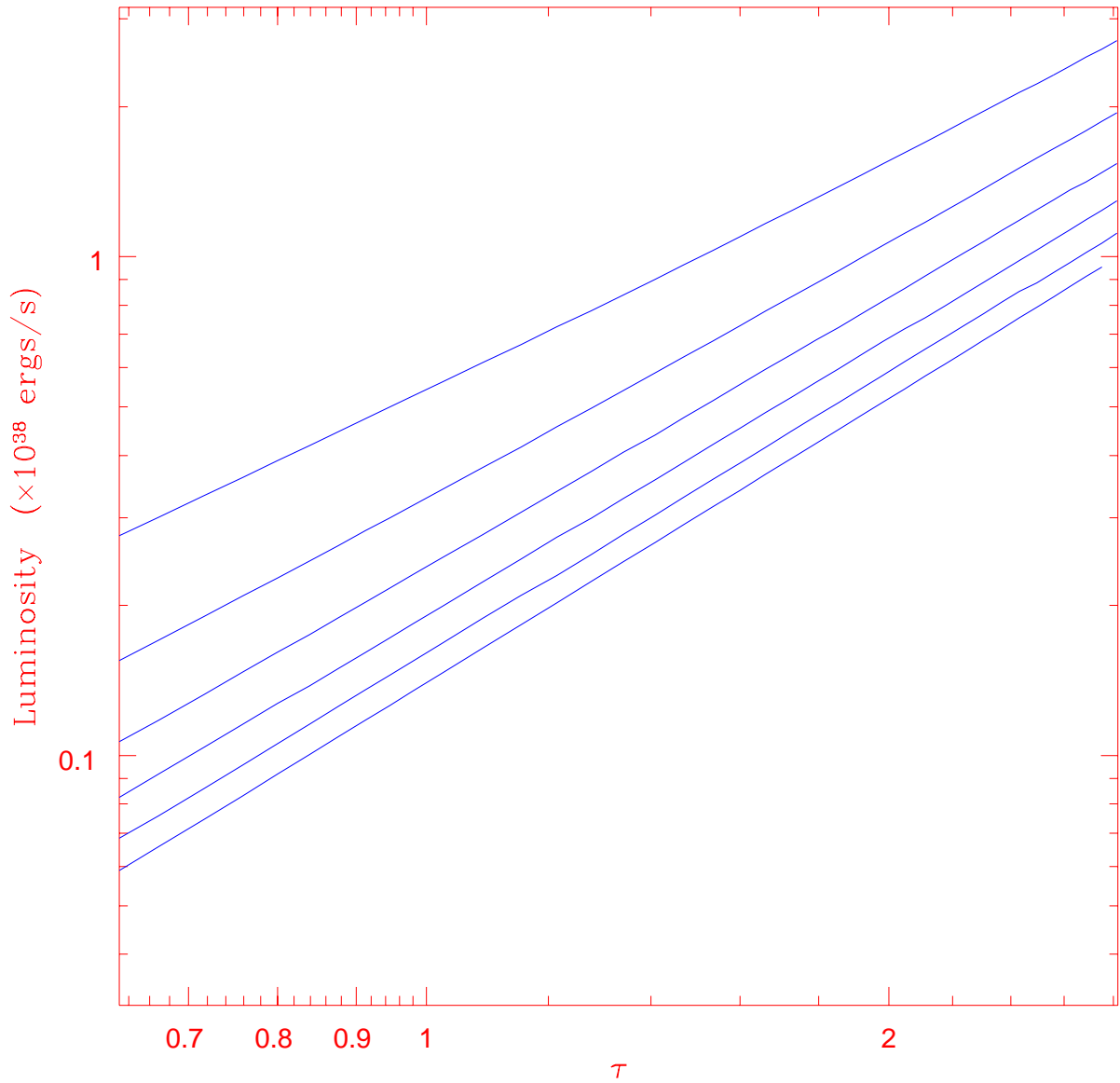}}}
\caption{\label{fig:3}Contours of mass outflow rate ($\dot M_o$) for luminosity and optical
depth. The contours correspond to from top to bottom $\dot M_o = 3 \times 10^{18},
10^{18}, 3 \times 10^{17}, 10^{17}, 3 \times 10^{16}, 10^{16}$ g/s. 
The other parameters are: $M = 10 M_\odot$, $R = 20 GM/c^2$, $\Gamma = 1.01$,
$kT_e = 50$ keV.}
\end{figure*}

\begin{figure*}
\hspace{-1.5cm} 
{\mbox{\psboxto(17cm;20cm){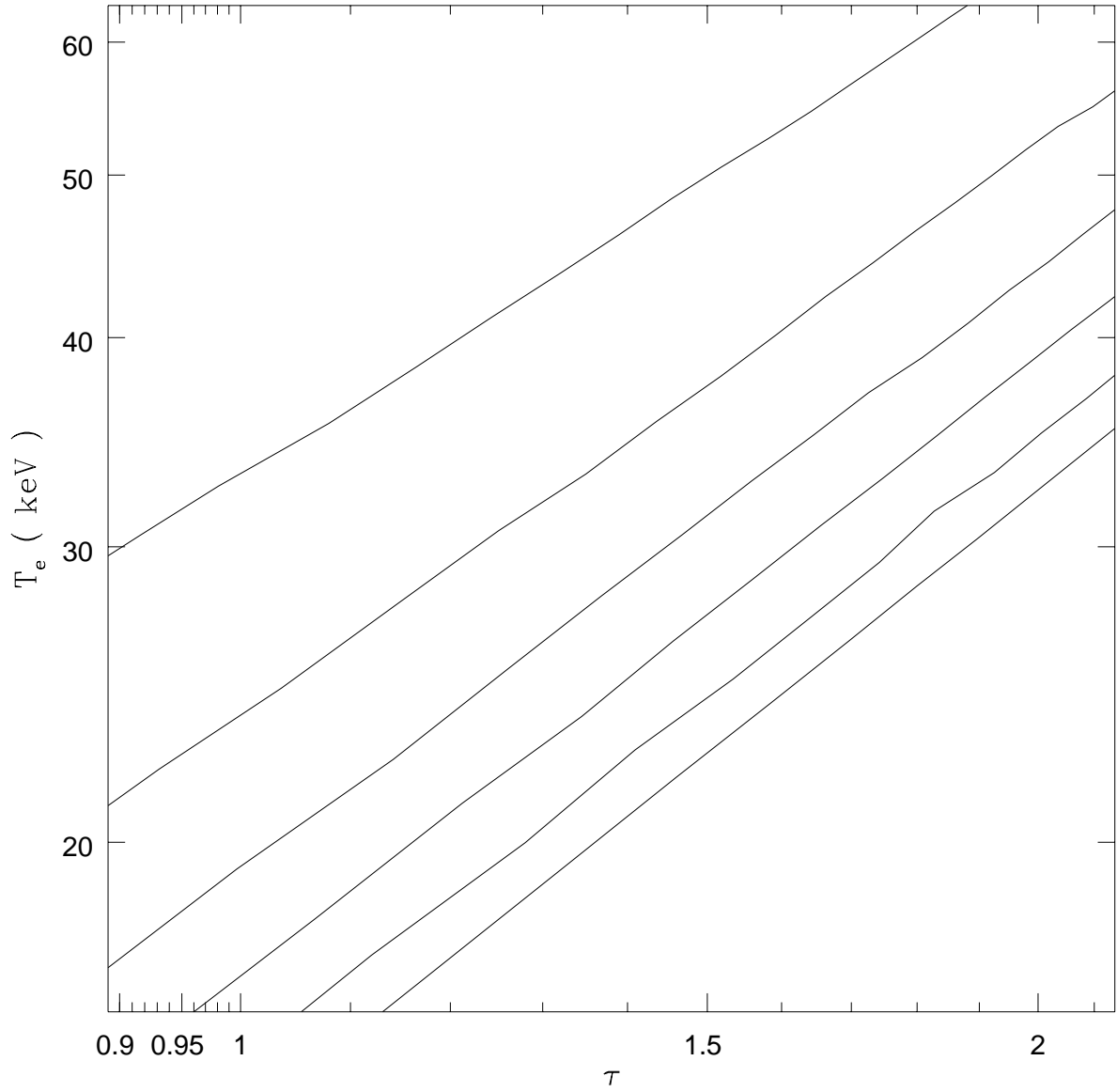}}}
\caption{\label{fig:4}Same as in Figure 3, except that the contours are for electron 
temperature and optical depth for a luminosity of $ L = 10^{38}$ ergs/s.}
\end{figure*}
\end{document}

%% file: psbox.tex
%
%
%
%
%
\def\temp{1.34}%
\let\tempp=\relax
\expandafter\ifx\csname psboxversion\endcsname\relax
  \message{PSBOX(\temp) loading}%
\else
    \ifdim\temp cm>\psboxversion cm
      \message{PSBOX(\temp) loading}%
    \else
      \message{PSBOX(\psboxversion) is already loaded: I won't load
        PSBOX(\temp)!}%
      \let\temp=\psboxversion
      \let\tempp= 
    \fi
\fi
\tempp
\let\psboxversion=\temp
\catcode`\@=11
%
%
\def\psfortextures{
\def\PSspeci@l##1##2{%
\special{illustration ##1\space scaled ##2}%
}}%
\def\psfordvitops{
\def\PSspeci@l##1##2{%
\special{dvitops: import ##1\space \the\drawingwd \the\drawinght}%
}}%
\def\psfordvips{
\def\PSspeci@l##1##2{%
\d@my=0.1bp \d@mx=\drawingwd \divide\d@mx by\d@my
\includegraphics{##1\space}}}%
\def\psforoztex{
\def\PSspeci@l##1##2{%
\special{##1 \space
      ##2 1000 div dup scale
      \number-\psllx\space \number-\pslly\space translate
}}}%
\def\psfordvitps{
\def\psdimt@n@sp##1{\d@mx=##1\relax\edef\psn@sp{\number\d@mx}}
\def\PSspeci@l##1##2{%
\special{dvitps: Include0 "psfig.psr"}
\psdimt@n@sp{\drawingwd}
\special{dvitps: Literal "\psn@sp\space"}
\psdimt@n@sp{\drawinght}
\special{dvitps: Literal "\psn@sp\space"}
\psdimt@n@sp{\psllx bp}
\special{dvitps: Literal "\psn@sp\space"}
\psdimt@n@sp{\pslly bp}
\special{dvitps: Literal "\psn@sp\space"}
\psdimt@n@sp{\psurx bp}
\special{dvitps: Literal "\psn@sp\space"}
\psdimt@n@sp{\psury bp}
\special{dvitps: Literal "\psn@sp\space startTexFig\space"}
\special{dvitps: Include1 "##1"}
\special{dvitps: Literal "endTexFig\space"}
}}%
\def\psfordvialw{
\def\PSspeci@l##1##2{
\special{language "PostScript",
position = "bottom left",
literal "  \psllx\space \pslly\space translate
  ##2 1000 div dup scale
  -\psllx\space -\pslly\space translate",
include "##1"}
}}%
\def\psforptips{
\def\PSspeci@l##1##2{{
\d@mx=\psurx bp
\advance \d@mx by -\psllx bp
\divide \d@mx by 1000\multiply\d@mx by \xscale
\incm{\d@mx}
\let\tmpx\dimincm
\d@my=\psury bp
\advance \d@my by -\pslly bp
\divide \d@my by 1000\multiply\d@my by \xscale
\incm{\d@my}
\let\tmpy\dimincm
\d@mx=-\psllx bp
\divide \d@mx by 1000\multiply\d@mx by \xscale
\d@my=-\pslly bp
\divide \d@my by 1000\multiply\d@my by \xscale
\at(\d@mx;\d@my){\special{ps:##1 x=\tmpx, y=\tmpy}}
}}}%
\def\psonlyboxes{
\def\PSspeci@l##1##2{%
\at(0cm;0cm){\boxit{\vbox to\drawinght
  {\vss\hbox to\drawingwd{\at(0cm;0cm){\hbox{({\tt##1})}}\hss}}}}
}}%
\def\psloc@lerr#1{%
\let\savedPSspeci@l=\PSspeci@l%
\def\PSspeci@l##1##2{%
\at(0cm;0cm){\boxit{\vbox to\drawinght
  {\vss\hbox to\drawingwd{\at(0cm;0cm){\hbox{({\tt##1}) #1}}\hss}}}}
\let\PSspeci@l=\savedPSspeci@l
}}%
%
%
\newread\pst@mpin
\newdimen\drawinght\newdimen\drawingwd
\newdimen\psxoffset\newdimen\psyoffset
\newbox\drawingBox
\newcount\xscale \newcount\yscale \newdimen\pscm\pscm=1cm
\newdimen\d@mx \newdimen\d@my
\newdimen\pswdincr \newdimen\pshtincr
\let\ps@nnotation=\relax
{\catcode`\|=0 |catcode`|\=12 |catcode`|
|catcode`#=12 |catcode`*=14
|xdef|backslashother{\}*
|xdef|percentother{
|xdef|tildeother{~}*
|xdef|sharpother{#}*
}%
\def\R@moveMeaningHeader#1:->{}%
\def\uncatcode#1{%
\edef#1{\expandafter\R@moveMeaningHeader\meaning#1}}%
\def\execute#1{#1}
\def\psm@keother#1{\catcode`#112\relax}
\def\executeinspecs#1{%
\execute{\begingroup\let\do\psm@keother\dospecials\catcode`\^^M=9#1\endgroup}}%
\def\@mpty{}%
\def\matchexpin#1#2{
  \fi%
  \edef\tmpb{{#2}}%
  \expandafter\makem@tchtmp\tmpb%
  \edef\tmpa{#1}\edef\tmpb{#2}%
  \expandafter\expandafter\expandafter\m@tchtmp\expandafter\tmpa\tmpb\endm@tch%
  \if\match%
}%
\def\matchin#1#2{%
  \fi%
  \makem@tchtmp{#2}%
  \m@tchtmp#1#2\endm@tch%
  \if\match%
}%
\def\makem@tchtmp#1{\def\m@tchtmp##1#1##2\endm@tch{%
  \def\tmpa{##1}\def\tmpb{##2}\let\m@tchtmp=\relax%
  \ifx\tmpb\@mpty\def\match{YN}%
  \else\def\match{YY}\fi%
}}%
\def\incm#1{{\psxoffset=1cm\d@my=#1
 \d@mx=\d@my
  \divide\d@mx by \psxoffset
  \xdef\dimincm{\number\d@mx.}
  \advance\d@my by -\number\d@mx cm
  \multiply\d@my by 100
 \d@mx=\d@my
  \divide\d@mx by \psxoffset
  \edef\dimincm{\dimincm\number\d@mx}
  \advance\d@my by -\number\d@mx cm
  \multiply\d@my by 100
 \d@mx=\d@my
  \divide\d@mx by \psxoffset
  \xdef\dimincm{\dimincm\number\d@mx}
}}%
%
\newif\ifNotB@undingBox
\newhelp\PShelp{Proceed: you'll have a 5cm square blank box instead of
your graphics (Jean Orloff).}%
\def\s@tsize#1 #2 #3 #4\@ndsize{
  \def\psllx{#1}\def\pslly{#2}%
  \def\psurx{#3}\def\psury{#4}
  \ifx\psurx\@mpty\NotB@undingBoxtrue
  \else
    \drawinght=#4bp\advance\drawinght by-#2bp
    \drawingwd=#3bp\advance\drawingwd by-#1bp
  \fi
  }%
\def\sc@nBBline#1:#2\@ndBBline{\edef\p@rameter{#1}\edef\v@lue{#2}}%
\def\g@bblefirstblank#1#2:{\ifx#1 \else#1\fi#2}%
{\catcode`\%=12
\xdef\B@undingBox{
\def\ReadPSize#1{
 \readfilename#1\relax
 \let\PSfilename=\lastreadfilename
 \openin\pst@mpin=#1\relax
 \ifeof\pst@mpin \errhelp=\PShelp
   \errmessage{I haven't found your postscript file (\PSfilename)}%
   \psloc@lerr{was not found}%
   \s@tsize 0 0 142 142\@ndsize
   \closein\pst@mpin
 \else
   \if\matchexpin{\GlobalInputList}{, \lastreadfilename}%
   \else\xdef\GlobalInputList{\GlobalInputList, \lastreadfilename}%
     \immediate\write\psbj@inaux{\lastreadfilename,}%
   \fi%
   \loop
     \executeinspecs{\catcode`\ =10\global\read\pst@mpin to\n@xtline}%
     \ifeof\pst@mpin
       \errhelp=\PShelp
       \errmessage{(\PSfilename) is not an Encapsulated PostScript File:
           I could not find any \B@undingBox: line.}%
       \edef\v@lue{0 0 142 142:}%
       \psloc@lerr{is not an EPSFile}%
       \NotB@undingBoxfalse
     \else
       \expandafter\sc@nBBline\n@xtline:\@ndBBline
       \ifx\p@rameter\B@undingBox\NotB@undingBoxfalse
         \edef\t@mp{%
           \expandafter\g@bblefirstblank\v@lue\space\space\space}%
         \expandafter\s@tsize\t@mp\@ndsize
       \else\NotB@undingBoxtrue
       \fi
     \fi
   \ifNotB@undingBox\repeat
   \closein\pst@mpin
 \fi
\message{#1}%
}%
%
%
\def\psboxto(#1;#2)#3{\vbox{
   \ReadPSize{#3}%
   \divide\drawingwd by 1000
   \divide\drawinght by 1000
   \d@mx=#1
   \ifdim\d@mx=0pt\xscale=1000
         \else \xscale=\d@mx \divide \xscale by \drawingwd\fi
   \d@my=#2
   \ifdim\d@my=0pt\yscale=1000
         \else \yscale=\d@my \divide \yscale by \drawinght\fi
   \ifnum\yscale=1000
         \else\ifnum\xscale=1000\xscale=\yscale
                    \else\ifnum\yscale<\xscale\xscale=\yscale\fi
              \fi
   \fi
   \divide\pswdincr by 1000 \multiply\pswdincr by \xscale
   \divide\pshtincr by 1000 \multiply\pshtincr by \xscale
   \divide\psxoffset by1000 \multiply\psxoffset by\xscale
   \divide\psyoffset by1000 \multiply\psyoffset by\xscale
   \global\divide\pscm by 1000
   \global\multiply\pscm by\xscale
   \multiply\drawingwd by\xscale \multiply\drawinght by\xscale
   \ifdim\d@mx=0pt\d@mx=\drawingwd\fi
   \ifdim\d@my=0pt\d@my=\drawinght\fi
   \message{scaled \the\xscale}%
 \hbox to\d@mx{\hss\vbox to\d@my{\vss
   \global\setbox\drawingBox=\hbox to 0pt{\kern\psxoffset\vbox to 0pt{
      \kern-\psyoffset
      \PSspeci@l{\PSfilename}{\the\xscale}%
      \vss}\hss\ps@nnotation}%
   \advance\pswdincr by \drawingwd
   \advance\pshtincr by \drawinght
   \global\wd\drawingBox=\the\pswdincr
   \global\ht\drawingBox=\the\pshtincr
   \baselineskip=0pt
   \copy\drawingBox
 \vss}\hss}%
  \global\psxoffset=0pt
  \global\psyoffset=0pt
  \global\pswdincr=0pt
  \global\pshtincr=0pt 
  \global\pscm=1cm 
  \global\drawingwd=\drawingwd
  \global\drawinght=\drawinght
}}%
%
%
\def\psboxscaled#1#2{\vbox{
  \ReadPSize{#2}%
  \xscale=#1
  \message{scaled \the\xscale}%
  \advance\drawingwd by\pswdincr\advance\drawinght by\pshtincr
  \divide\pswdincr by 1000 \multiply\pswdincr by \xscale
  \divide\pshtincr by 1000 \multiply\pshtincr by \xscale
  \divide\psxoffset by1000 \multiply\psxoffset by\xscale
  \divide\psyoffset by1000 \multiply\psyoffset by\xscale
  \divide\drawingwd by1000 \multiply\drawingwd by\xscale
  \divide\drawinght by1000 \multiply\drawinght by\xscale
  \global\divide\pscm by 1000
  \global\multiply\pscm by\xscale
  \global\setbox\drawingBox=\hbox to 0pt{\kern\psxoffset\vbox to 0pt{
     \kern-\psyoffset
     \PSspeci@l{\PSfilename}{\the\xscale}%
     \vss}\hss\ps@nnotation}%
  \advance\pswdincr by \drawingwd
  \advance\pshtincr by \drawinght
  \global\wd\drawingBox=\the\pswdincr
  \global\ht\drawingBox=\the\pshtincr
  \baselineskip=0pt
  \copy\drawingBox
  \global\psxoffset=0pt
  \global\psyoffset=0pt
  \global\pswdincr=0pt
  \global\pshtincr=0pt 
  \global\pscm=1cm
  \global\drawingwd=\drawingwd
  \global\drawinght=\drawinght
}}%
%
\def\psbox#1{\psboxscaled{1000}{#1}}%
\newif\ifn@teof\n@teoftrue
\newif\ifc@ntrolline
\newif\ifmatch
\newread\j@insplitin
\newwrite\j@insplitout
\newwrite\psbj@inaux
\immediate\openout\psbj@inaux=psbjoin.aux
\immediate\write\psbj@inaux{\string\joinfiles}%
\immediate\write\psbj@inaux{\jobname,}%
%
%
\def\toother#1{\ifcat\relax#1\else\expandafter%
  \toother@ux\meaning#1\endtoother@ux\fi}%
\def\toother@ux#1 #2#3\endtoother@ux{\def\tmp{#3}%
  \ifx\tmp\@mpty\def\tmp{#2}\let\next=\relax%
  \else\def\next{\toother@ux#2#3\endtoother@ux}\fi%
\next}%
%
%
\let\readfilenamehook=\relax
\def\re@d{\expandafter\re@daux}
\def\re@daux{\futurelet\nextchar\stopre@dtest}%
\def\re@dnext{\xdef\lastreadfilename{\lastreadfilename\nextchar}%
  \afterassignment\re@d\let\nextchar}%
\def\stopre@d{\egroup\readfilenamehook}%
\def\stopre@dtest{%
  \ifcat\nextchar\relax\let\nextread\stopre@d
  \else
    \ifcat\nextchar\space\def\nextread{%
      \afterassignment\stopre@d\chardef\nextchar=`}%
    \else\let\nextread=\re@dnext
      \toother\nextchar
      \edef\nextchar{\tmp}%
    \fi
  \fi\nextread}%
\def\readfilename{\vbox\bgroup%
  \let\\=\backslashother \let\%=\percentother \let\~=\tildeother
  \let\#=\sharpother \xdef\lastreadfilename{}%
  \re@d}%
%
%
\xdef\GlobalInputList{\jobname}%
\def\psnewinput{%
  \def\readfilenamehook{
    \if\matchexpin{\GlobalInputList}{, \lastreadfilename}%
    \else\xdef\GlobalInputList{\GlobalInputList, \lastreadfilename}%
      \immediate\write\psbj@inaux{\lastreadfilename,}%
    \fi%
    \ps@ldinput\lastreadfilename\relax%
    \let\readfilenamehook=\relax%
  }\readfilename%
}%
\expandafter\ifx\csname @@input\endcsname\relax    
  \immediate\let\ps@ldinput=\input\def\input{\psnewinput}%
\else
  \immediate\let\ps@ldinput=\@@input
  \def\@@input{\psnewinput}%
\fi%
\def\nowarnopenout{%
 \def\warnopenout##1##2{%
   \readfilename##2\relax
   \message{\lastreadfilename}%
   \immediate\openout##1=\lastreadfilename\relax}}%
\def\warnopenout#1#2{%
 \readfilename#2\relax
 \def\t@mp{TrashMe,psbjoin.aux,psbjoint.tex,}\uncatcode\t@mp
 \if\matchexpin{\t@mp}{\lastreadfilename,}%
 \else
   \immediate\openin\pst@mpin=\lastreadfilename\relax
   \ifeof\pst@mpin
     \else
     \errhelp{If the content of this file is so precious to you, abort (ie
press x or e) and rename it before retrying.}%
     \errmessage{I'm just about to replace your file named \lastreadfilename}%
   \fi
   \immediate\closein\pst@mpin
 \fi
 \message{\lastreadfilename}%
 \immediate\openout#1=\lastreadfilename\relax}%
{\catcode`\%=12\catcode`\*=14
\gdef\splitfile#1{*
 \readfilename#1\relax
 \immediate\openin\j@insplitin=\lastreadfilename\relax
 \ifeof\j@insplitin
   \message{! I couldn't find and split \lastreadfilename!}*
 \else
   \immediate\openout\j@insplitout=TrashMe
   \message{< Splitting \lastreadfilename\space into}*
   \loop
     \ifeof\j@insplitin
       \immediate\closein\j@insplitin\n@teoffalse
     \else
       \n@teoftrue
       \executeinspecs{\global\read\j@insplitin to\spl@tinline\expandafter
         \ch@ckbeginnewfile\spl@tinline
       \ifc@ntrolline
       \else
         \toks0=\expandafter{\spl@tinline}*
         \immediate\write\j@insplitout{\the\toks0}*
       \fi
     \fi
   \ifn@teof\repeat
   \immediate\closeout\j@insplitout
 \fi\message{>}*
}*
\gdef\ch@ckbeginnewfile#1
 \def\t@mp{#1}*
 \ifx\@mpty\t@mp
   \def\t@mp{#3}*
   \ifx\@mpty\t@mp
     \global\c@ntrollinefalse
   \else
     \immediate\closeout\j@insplitout
     \warnopenout\j@insplitout{#2}*
     \global\c@ntrollinetrue
   \fi
 \else
   \global\c@ntrollinefalse
 \fi}*
\gdef\joinfiles#1\into#2{*
 \message{< Joining following files into}*
 \warnopenout\j@insplitout{#2}*
 \message{:}*
 {*
 \edef\w@##1{\immediate\write\j@insplitout{##1}}*
\w@{
\w@{
\w@{
\w@{
\w@{
\w@{
\w@{
\w@{
\w@{
\w@{
\w@{\string\input\space psbox.tex}*
\w@{\string\splitfile{\string\jobname}}*
\w@{\string\let\string\autojoin=\string\relax}*
}*
 \expandafter\tre@tfilelist#1, \endtre@t
 \immediate\closeout\j@insplitout
 \message{>}*
}*
\gdef\tre@tfilelist#1, #2\endtre@t{*
 \readfilename#1\relax
 \ifx\@mpty\lastreadfilename
 \else
   \immediate\openin\j@insplitin=\lastreadfilename\relax
   \ifeof\j@insplitin
     \errmessage{I couldn't find file \lastreadfilename}*
   \else
     \message{\lastreadfilename}*
     \immediate\write\j@insplitout{
     \executeinspecs{\global\read\j@insplitin to\oldj@ininline}*
     \loop
       \ifeof\j@insplitin\immediate\closein\j@insplitin\n@teoffalse
       \else\n@teoftrue
         \executeinspecs{\global\read\j@insplitin to\j@ininline}*
         \toks0=\expandafter{\oldj@ininline}*
         \let\oldj@ininline=\j@ininline
         \immediate\write\j@insplitout{\the\toks0}*
       \fi
     \ifn@teof
     \repeat
   \immediate\closein\j@insplitin
   \fi
   \tre@tfilelist#2, \endtre@t
 \fi}*
}%
\def\autojoin{%
 \immediate\write\psbj@inaux{\string\into{psbjoint.tex}}%
 \immediate\closeout\psbj@inaux
 \expandafter\joinfiles\GlobalInputList\into{psbjoint.tex}%
}%
%
%
%
\def\centinsert#1{\midinsert\line{\hss#1\hss}\endinsert}%
\def\psannotate#1#2{\vbox{%
  \def\ps@nnotation{#2\global\let\ps@nnotation=\relax}#1}}%
\def\pscaption#1#2{\vbox{%
   \setbox\drawingBox=#1
   \copy\drawingBox
   \vskip\baselineskip
   \vbox{\hsize=\wd\drawingBox\setbox0=\hbox{#2}%
     \ifdim\wd0>\hsize
       \noindent\unhbox0\tolerance=5000
    \else\centerline{\box0}%
    \fi
}}}%
%
\def\at(#1;#2)#3{\setbox0=\hbox{#3}\ht0=0pt\dp0=0pt
  \rlap{\kern#1\vbox to0pt{\kern-#2\box0\vss}}}%
%
\newdimen\gridht \newdimen\gridwd
\def\gridfill(#1;#2){%
  \setbox0=\hbox to 1\pscm
  {\vrule height1\pscm width.4pt\leaders\hrule\hfill}%
  \gridht=#1
  \divide\gridht by \ht0
  \multiply\gridht by \ht0
  \gridwd=#2
  \divide\gridwd by \wd0
  \multiply\gridwd by \wd0
  \advance \gridwd by \wd0
  \vbox to \gridht{\leaders\hbox to\gridwd{\leaders\box0\hfill}\vfill}}%
%
\def\fillinggrid{\at(0cm;0cm){\vbox{%
  \gridfill(\drawinght;\drawingwd)}}}%
%
%
\def\textleftof#1:{%
  \setbox1=#1
  \setbox0=\vbox\bgroup
    \advance\hsize by -\wd1 \advance\hsize by -2em}%
\def\textrightof#1:{%
  \setbox0=#1
  \setbox1=\vbox\bgroup
    \advance\hsize by -\wd0 \advance\hsize by -2em}%
\def\endtext{%
  \egroup
  \hbox to \hsize{\valign{\vfil##\vfil\cr%
\box0\cr%
\noalign{\hss}\box1\cr}}}%
%
\def\frameit#1#2#3{\hbox{\vrule width#1\vbox{%
  \hrule height#1\vskip#2\hbox{\hskip#2\vbox{#3}\hskip#2}%
        \vskip#2\hrule height#1}\vrule width#1}}%
\def\boxit#1{\frameit{0.4pt}{0pt}{#1}}%
\catcode`\@=12 
%
 \psfordvips   